# Band-edge Exciton Fine Structure and Exciton Recombination Dynamics in Single crystals of Layered Hybrid Perovskites

Hong-Hua Fang,[1] Jie Yang,[1] Sampson Adjokatse,[1] Machteld E. Kamminga,[1] Herman Duim,[1] Jianting Ye,[1] Graeme R. Blake,[1] Jacky Even,[2] and Maria Antonietta Loi[1]*

[1]Zernike Institute for Advanced Materials, University of Groningen, Nijenborgh 4, 9747 AG, Groningen, The Netherlands
[2]Univ Rennes, INSA Rennes, CNRS, Institut FOTON - UMR 6082, F-35000 Rennes, France
Corresponding Author: m.a.loi@rug.nl

Two-dimensional (2D) perovskite materials have recently re-attracted intense research interest for applications in photovoltaics and optoelectronics. As a consequence of the dielectric and quantum confinement effect, they show strongly bound and stable excitons at room temperature. In this report, the band-edge exciton fine structure and in particular its exciton and biexciton dynamics in high quality crystals of $(PEA)_2PbI_4$ are investigated. A comparison of bulk and surface exciton lifetimes yields a room temperature surface recombination velocity of $2\times10^3$ cm/s and an intrinsic lifetime of 185 ns. Biexciton emission is evidenced at room temperature, with binding energy of ≈ 45 meV and a lifetime of 80 ps. At low temperature, exciton state splitting is observed, which is caused by the electron-hole exchange interaction. Transient photoluminescence resolves the low-lying dark exciton state, with a bright/dark splitting energy estimated to be 10 meV. This work contributes to understand the complex scenario of the elementary photoexcitations in 2D perovskites.

Hybrid organic-inorganic perovskites are currently under the spotlight for optoelectronic applications due to their remarkable photophysical properties and solution processability.[1] They have been used to demonstrate highly efficient solar cells,[2,3] light emitting diodes,[4,5] and gas sensors.[6,7] Very recently, quasi-2D Ruddlesden–Popper perovskites, a family of layered compounds with tunable semiconductor characteristics, have also been explored for optoelectronic devices.[8–11] The thickness (n) of the perovskite sheets can be synthetically controlled by adjusting the ratio between the spacer cation and the small organic cation, thus allowing the onset of absorption to be tuned from violet to near infrared.[12,13] Furthermore, layered perovskites afford specific advantages over other inorganic 2D materials such as transition metal dichalcogenides (TMDCs), phosphorene, and graphene. They can be easily grown by both solution methods and vapor transport methods at low temperature[14–17], with a tunable bulk direct bandgap.[18] These advantages make them appealing for future optoelectronic and photonic applications.

Unlike their 3D counterparts, the dielectric and quantum confinement of carriers in the 2D perovskite layers gives rise to unusually strong excitonic effects.[19,20] It has been experimentally observed that excitons are tightly confined in the inorganic layers with binding energy as high as a few hundred millielectronvolts (significantly higher than that of 3D perovskites).[21] This greatly enhanced exciton binding energy makes them particularly interesting for light-emitting applications.[8,22] Moreover, 2D perovskites can exhibit a variety of multiexciton species, including biexcitons and trions.[23–25] The presence of these quasiparticles is exciting due to their unique role in leading to a better understanding of many body effects and their great promise for photonic applications. In addition, recent experiments reveal an important role of electron–phonon couplings on the exciton dynamics in two-dimensional lead-iodide perovskite, suggesting a complex scenario for carrier relaxation and exciton formation.[26,27] It is therefore crucial to understand elementary photoexcitations in these layered materials. However, exciton fine structures and their properties are usually masked by local energy fluctuations resulting from disorder in thin films or broad emission due to the formation of self-trapped excitons.[28] Whereas their steady-state optical properties have been intensively investigated,[29–33] the investigation of exciton-exciton interactions, biexcitons and their recombination dynamics remains modest.[34,35]

In this work, temperature- and power-dependent time-resolved photoluminescence are performed to investigate band-edge excitonic features of two-dimensional perovskite. Mechanically exfoliated (PEA)$_2$PbI$_4$ crystals instead of thin films are employed for studying the excitonic fine structure. These perovskites show longer exciton lifetimes of $\tau_1$ = 185 ns and $\tau_2$ = 1.19 μs, yielding better access to the intrinsic properties of layered halide perovskites than thin films.[36,37] Biexciton emission is demonstrated at room temperature, with a lifetime of 80 ps. Along biexciton binding energies of up to 45 meV is observed. Exciton state fine structure at low temperature are resolved, with a bright/dark splitting energy estimated to be 10 meV. This enables us understanding the complex scenario for carrier relaxation and exciton formation, and the interplay between biexciton and exciton relaxation.

Figure 1(a) shows a bulk crystal synthesized by the anti-solvent vapor-assisted crystallization (AVC) method. The X-ray diffraction (XRD) pattern of a flake exfoliated from a (PEA)$_2$PbI$_4$ bulk crystal is shown in Figure S1 of the Supporting Information. The well-defined diffraction peaks correspond to the (*00l*) series of reflections, indicating good crystallinity with the layers stacking perpendicular to the surface of the flake. The crystal structure determined from X-ray diffraction data collected at 100 K on a single crystal is displayed in Figure 1(b). The unit cell ($a$ = 6.1594(4) Å, $b$ = 6.0991(4) Å, $c$ = 32.261(2) Å, $\beta$ = 94.299(2)° at 100 K) is similar to that of the monoclinic *C*2/*m* structure reported by Calabrese et al.[38] (if the *a*- and *c*-axes are swapped), but many weak peaks violating the reflection condition *hkl*, *k*+*l*=2*n* for A-centering (C-centering in the unit cell setting of Calabrese et al.) were observed (Figure 1(c)). Using our unit cell setting, the conditions *h*0*l*, *l*=2*n* and 0*k*0, *k*=2*n* strictly hold, which implies that the space group is *P*2$_1$/*c*. Structure solution by direct methods and subsequent refinement showed that the PbI$_6$ octahedra are disordered over two orientations by a rotation of 28.5° around the long *c*-axis (Figure S3 in Supporting Information), in similar fashion to the structure reported by Calabrese et al.[38] The structural model was refined using a single PEA molecule, but with N atoms of the ammonium group also disordered over two positions. The large ADPs (atomic displacement parameters) of the carbon atoms in the phenyl ring suggest that there is a degree of disorder in the position of the entire PEA molecule. No phase transition was detected in the temperature range 100 – 295 K (Table S1, in Supporting Information).

The 2D perovskite crystals can be mechanically exfoliated using the scotch tape method. Figure 1(d) and (e) show an optical microscopy image and the corresponding photoluminescence image (obtained with a 488 nm laser diode) of multiple mechanically exfoliated (PEA)$_2$PbI$_4$ crystals, respectively. The homogenous surface indicates a good quality of the exfoliated crystal. The edges of the largest crystal (and some cracks on its surface) show stronger PL intensity compared to the inner region, indicating the waveguide effect in these layered perovskites. Atomic force microscopy (AFM) working in tapping mode was used to characterize the surface of the crystals, revealing step-like morphology (Figure 1(f)). The average step height generated from the topographic image is ~1.65 nm, which is consistent with the distance between the inorganic layers along the *c*-axis of the crystal structure (a single unit cell (i.e., a bilayer) of the bulk material is ~3.2 nm in thickness).

Freshly cleaved crystals show a quantum yield as high as 86 %. In our previous report, we show that surface traps can be introduced by illumination, which strongly affects the photoexcitation recombination dynamics.[39] To discern the surface and bulk recombination kinetics, one- and two-photon-excited photoluminescence in single crystals are comparatively investigated. The absorption spectrum of thin films (Figure S4 in Supporting Information) exhibits a very strong excitonic absorption at a photon energy of 515 nm (2.4eV). The absorption depth in (PEA)$_2$PbI$_4$ is estimated to be about 100 nm according to the absorption coefficient measured in the thin films, suggesting that photocarriers are located within the near surface region when under one-photon excitation. Therefore, it is expected that the exciton recombination under one-photon excitation is highly sensitive to the surface traps.[7] Under two-photon excitation, most of the carriers are created within the interior of the crystal; in this case, the exciton recombination dynamics should primarily reflect the bulk properties.[40] Therefore, we can distinguish between the surface and bulk contributions. The time-integrated one-photon (3.1 eV, 400 nm) and two-photon-pumped (1.55 eV, 800 nm) PL spectra of a (PEA)$_2$PbI$_4$ single crystal are presented in Figure 2. The one photon-excited PL shows a single emission band which peaks at 528 nm (2.35eV) (Figure 2-B), while the two-photon-pumped PL is redshifted to 546 nm (2.27eV), which is likely caused by reabsorption of the absorption tail of the material. This is due to the fact that the two-photon excitation has a much longer optical penetration depth (it can extend to be more than 100 μm), and the emission within the interior region of the crystal has to travel a long distance before it can escape to the surface.

The bulk recombination lifetime $\tau_b$ was determined from the TRPL measured by two-photon excitation, as shown in Figure 2(F). The curve can be fitted with biexponential decay, with lifetime $\tau_1$ = 185 ns, and $\tau_2$ = 1.19 μs. The bulk recombination rate is estimated to be $1/\tau_1$ = $5.4 \times 10^6$ s$^{-1}$, using the fast component, which is much smaller than that reported in two-dimensional (2D) transition metal dichalcogenides. The lifetime under one-photon excitation is $\tau_1$ = 2.5 ns, and $\tau_2$ = 30 ns. We attribute the faster recombination to the presence of surface traps. From these data, the surface recombination velocity (SRV) can be quantified by using the following equation:[41] $1/\tau_{eff}=1/\tau_b+\alpha S$, where $\alpha$ is the absorption coefficient at the wavelength of excitation. The SRV is estimated to be $2.0 \times 10^3$ cm/s, which is comparable to the values reported for the 3D perovskite counterparts.[7,42] We note that previous measurements on thin films yield an even shorter lifetime of 0.64ns at room temperature,[37] also showing that carrier recombination is dominated by surface effects.

Since the optical properties of low dimensional materials are usually dominated by excitons, it is important to understand their behavior at varied densities, especially at higher excitation, where exciton-exciton interactions cannot be ignored, as schematically shown in Figure 3(A). We measured photoluminescence spectra of $(PEA)_2PbI_4$ under elevated optical excitation with a 3.1 eV (400 nm) femtosecond laser pulse. Representative PL spectra of the $(PEA)_2PbI_4$ crystal as a function of excitation fluence are shown in Figure 3(B). We note the emergence of an additional PL shoulder at high-power excitation, suggesting the formation of complex excitonic states at higher exciton density. The integrated PL intensity is plotted against the excitation fluence in Figure 3(C). The PL intensity is linearly dependent on the excitation intensity for pump fluence $P < 0.5$ μJ/cm$_2$. This confirms that in this case the recombination process involves a single electron-hole pair. At higher excitation fluence, the integrated intensity of this crystal shows a sublinear increase as a function of the excitation power, with a clear saturation of the intensity for $P > 5$ μJ/cm$_2$. Considering that the additional PL shoulder emerges at high-power excitation, we speculate that this PL shoulder arises from radiative biexciton (XX), four-particle excitations consisting of two electrons and two holes bound together through Coulombic forces and exchange interactions, as shown in Figure 3(A). To isolate the new emission contribution, we evaluate the difference between the two spectra (see Figure S5, Supporting Information), from which we extract an energy difference of ΔE ≈ 45 meV. We note that in previous studies biexciton luminescence was only evidenced as broad emission below 100K in $(PEA)_2PbI_4$ thin films,[24] or below 16K in thin films based on other single-layer iodide perovskite compounds.[43,44]

To investigate the exciton interactions and study the room temperature biexciton emission, we performed ultrafast photoluminescence spectroscopy at varying excitation density. Figure 4 presents the time-resolved PL measurements for a (PEA)$_2$PbI$_4$ crystal under excitation of a 3.1 eV (400 nm) femtosecond laser. Figure 4(A) and 4(B) show spectral- and time-resolved PL images measured for excitations of 0.17 and 8.04 µJ/cm$_2$, respectively. The spectra in Figure 4(C) were obtained by extracting the corresponding transient PL spectra at t = 50 ps. Similarly to what was observed for the time-integrated PL spectra (Figure 3-B), a broad emission with an additional emission shoulder at low energy is clearly revealed under high fluence. The excitation power-dependent PL dynamics of the (PEA)$_2$PbI$_4$ crystal is illustrated in Figure 4(D). The PL decays are normalized to match their long-term decay. At low excitation fluence (< 0.2 µJ/cm$_2$) the PL decay curves are identical, suggesting that no additional exciton complex is generated in this case. With increasing fluence, an additional fast component appears, which is a typical signature of the formation of biexcitonic complexes due to a fast non-radiative "Auger"-like process.[45]

To validate that the fast components come from bi-exciton recombination, the signals were isolated from their long delay components. As shown in Figure 4(D), the long delay components, which are attributed to single exciton recombination, are identical at different excitation powers. Thus, the overall PL decay can be well described by a superposition of the long delay components $f(t)$ and an additional fast component, $A\exp(-t/\tau_{xx}) + Bf(t)$. The PL intensity from the fast-decay component can be obtained by subtracting the PL amplitude related to the slow-decay component at an early time after photoexcitation. Applying this, we obtain the fast recombination curves for different excitation powers, as shown in Figure 4(E). The lifetime for this fast decay at early time can easily be estimated as ~80 ps. Figure 4(F) presents the power dependence of the fast-decay component, as well as the single exciton contribution. The amplitude of the fast decay component in the range of low excitation power approximately scales as $P_2$, providing strong evidence that the fast component is associated with radiative recombination of biexcitons.

Biexciton binding energies are generally small, thus they can be efficiently thermalized at high temperature, making them unstable at room temperature. They can be stable only when the binding energy is higher than the thermal energy, $k_BT$ = 26 meV, where $k_B$ and T are the Boltzmann constant and room temperature, respectively. As mentioned before, at room temperature we only observed an additional PL shoulder without fine excitonic structure due to the strong thermal broadening of the exciton line. To further clarify exciton-biexciton recombination in the crystals, time-resolved PL spectra of single crystals under excitation of 400 nm were measured at low temperature. Additional peaks are clearly present at temperature T < 100 K in the time-integrated PL spectra (See Figure S6, supporting information). Figure 5(A) presents spectral- and time-resolved PL images of a $(PEA)_2PbI_4$ crystal measured at 5.4 K, under excitation of 0.23 μJ/cm$^2$ (left) and 2.3 μJ/cm$^2$ (right). The early-time (at $t$ = 70 ps) PL spectra of the $(PEA)_2PbI_4$ crystal are displayed in Figure 5(B). Unlike at room temperature, two distinct excitonic emission peaks are apparent: the first is located at around 2.351 eV, and the second at 2.309 eV. The inset shows the intensity of the 2.351 eV and 2.309 eV emission lines as a function of excitation density. When the pump intensity is increased, the peak at about 2.351 eV shows a dependence of $I = P^{0.84}$ on the excitation fluence (P), and the peak at 2.309 eV grows superlinearly (exponent of $k$ = 1.3) with the excitation power. We attribute it to the biexciton recombination (XX), whose density is predicted to scale quadratically with the exciton population. The reason for the deviation from k = 2 is possibly because of the lack of equilibrium between the states, as typically observed in quantum-well systems.[46,47] Photobleaching at higher powers may also be responsible for this behavior.

The biexciton binding energy is defined as the difference in energy between two free excitons and the biexciton state: $\Delta E_{XX} = 2E_X - E_{XX}$.[48] If we assume that biexciton radiative recombination gives a photon with energy $\hbar\omega_{XX}$ and leaves an exciton behind, XX → X +

photon, then $E_{xx} = \hbar\omega_{xx} + E_x = \hbar\omega_{xx} + \hbar\omega_x$, where $\hbar\omega_x$ denotes the exciton emission energy. Thus, the biexciton binding energy is given by the energy difference between these two transitions. From the emission peaks, we can obtain $\Delta E_{xx} = \hbar\omega_x - \hbar\omega_{xx} = 44$ meV, which is very close to the biexciton binding energy measured in other layered perovskites [29] and consistent with the broad emission line observed for (PEA)$_2$PbI$_4$ thin films.[24,29] This value is consistent with the energy difference of 45 meV measured at room temperature, supporting the notion that the broadened emission at high excitation is associated with the biexciton emergence at room temperature. This large biexciton binding energy could be expected because the excitons in (PEA)$_2$PbI$_4$ thin films have large binding energy, with measured values of around 200 meV from absorption spectra at low temperature (Figure S7 in Supporting Information). The large biexciton binding energies share the same origin as the one of the excitons, where both quantum and dielectric confinement greatly enhance the Coulomb interaction in these 2D structures. The large biexciton binding energy implies a large stokes shift in excitonic absorption, which could thus circumvent linear absorption losses. In this context, it is expected to help reducing the lasing threshold of these materials, as recently reported in two-dimensional CdSe colloidal nanosheets.[49] By comparing the exciton and biexciton binding energies, we can extract a Haynes factor of 0.2, which is similar to that found in quantum-wells.[50]

In order to assess the recombination dynamics, the PL spectra at different decay time after photoexcitation are depicted in Figure 5(C). Just after the excitation (t=70ps) at 400nm, the exciton is the predominant emission, while the XX peak intensity is comparable to the X at 1500 ps. This is further elucidated by the time evolution of the PL spectra, reported in Figure 5 (D). Decay curves at 2.355 and 2.309 eV show nonexponential decay. Fitting of the experimental curves in Figure 5(D) yields at 2.355 eV two decay times $t_1 = 39$ps and $t_2 = 253$ ps, and at 2.309 eV (XX emission) decay times of $t_1 = 51$ ps, $t_2 = 347$ ps. This is in contrast to the case at room temperature, where the biexciton lifetime is much shorter than the exciton lifetime. Generally, the decay time of the biexciton is controlled by the interplay between its formation and dissociation. At thermal equilibrium, the interconversion time between excitons and biexcitons is much shorter than the recombination lifetime, it is expected that the biexciton lifetime is about half that of the exciton, and that a quadratic relation exists between the density of biexcitons and excitons. However, this is not the case at low temperature for $(PEA)_2PbI_4$ single crystals. We note that the high energy peak red shifts from 2.355 eV to 2.342 eV as time elapses (Figure 5-A) and the TRPL curve at 2.342 eV shows a clear plateau (Figure 5-D), suggesting a population of excitons from higher energy states. The fitting of the PL dynamics at 2.342 eV clearly presents a rise time of 75 ps, which is of the same order as the fast decay time measured at 2.355 eV. We therefore postulate that an intermediate energy state lies at around 2.342 eV, which is 10 meV below the X state (2.355 eV), and that it plays an important role in the exciton-biexciton population dynamics.

We now discuss two possible origins for the state lying between the emission energies that we have attributed to the XX (2.309eV) and X (2.355eV) lines. The first possibility is the formation of trions, *i.e.* charged excitons, and the second to dark excitons. It has been reported that trions are efficiently generated only in 0D perovskite nanocrystals.[45,51–53] However, they usually suffer from rapid non-radiative Auger recombination, which exhibits faster decay than excitons. The hypothesis of trions can be therefore excluded by the longer relaxation time of the lower-lying state. Dark excitons (DE) are more likely responsible for the observed lower-lying state, which are ~10 meV below the emissive band-edge of the bright exciton (BE). This is further supported by the observation of strong quenching of light emission at reduced temperatures. With decreasing temperature, the thermalized exciton in the upper-level get trapped by the lower-lying state. Figure S8 in the supporting information displays the decrease in the exciton PL intensity observed when cooling from 150 K to 30 K. The decreases of PL intensity can be fitted by a very simple two-level model, including a high energy exciton state which is bright and a lower energy state, with an energy separation $\Delta E$. The red curve is obtain from this model with the parameter $\Delta E = 10$ meV, verifying the hypothesis of dark exciton. The formation of the dark state can be understood in terms of a splitting effect by the exchange interactions between the electron and the hole, leading to three fine-structure levels (optically allowed $\Gamma_5^-$ and optically forbidden $\Gamma_1^-$ and $\Gamma_2^-$ levels; the spectral resolution is not sufficient to resolve the $\Gamma_1^-$ and $\Gamma_2^-$ levels, therefore these excitons are referred to as $\Gamma_{1,2}^-$), as reported for $(C_4H_9NH_3)_2PbBr_4$.[31,54] Due to a fast spin relaxation from the $\Gamma_5^-$ to $\Gamma_{1,2}^-$ levels, the emission $\Gamma_5^-$ decays rapidly with a time constant of ~75 ps. In the meantime, the exciton population of the $\Gamma_{1,2}^-$ state increases, then decays with a much slower time constant of 300 ps, as shown in Figure 5(D). The low-lying states may also be a reason for the long radiative lifetime measured at room temperature. The spin splitting is relatively modest, with a size comparable to the room-temperature thermal energy, resulting in an efficient transition from the dark to the bright state, and facilitating light emission as thermally activated delayed luminescence.

   In summary, the temperature-dependent optical properties of single crystals of the layered perovskite $(PEA)_2PbI_4$ have been investigated. These measurements allow to resolve its band-edge exciton fine structure. Biexciton emission is observed at room temperature, with a binding energy of ~45 meV, and a lifetime of 80 ps. Such high biexciton binding energy may enable the use of these materials for lasing applications. In high quality single crystals, the splitting of the exciton is observed at low temperature, with a dark state/bright state energy splitting estimated to be 10 meV. The exciton fine structure plays an important role in the coupled exciton/biexciton population dynamics. These findings not only shed light on the understanding of the electron-hole correlations in layered perovskite systems, but also provide information for improving the performance of optoelectronic devices and bring them closer to potential applications in quantum information processing.

**Experimental Section**

*Materials:* Bulk crystals of (PEA)$_2$PbI$_4$ were synthesized by the anti-solvent vapor-assisted crystallization (AVC) method. (PEA)$_2$PbI$_4$ precursor solutions were prepared by dissolving PbI$_2$ and C$_6$H$_5$C$_2$H$_4$NH$_3$I (called PEAI hereafter) in Dimethylformamide (DMF) (1:2 molar ratio). 1 mol L$^{-1}$ solution of is poured into a small vial and then placed in a bigger Teflon cap vial containing the antisolvent: dichloromethane (DCM). After 48 hours, millimeter-sized rectangle-shaped orange crystals start to grow in the small vial. To grow thin films, the (PEA)$_2$PbI$_4$ precursor solutions were prepared by dissolving PEAI and PbI$_2$ with a molar ratio of 2:1 in a mixed solvent of DMF and dimethyl sulfoxide (DMSO) in a volume ratio of 4:1. The perovskite solution was spin-coated on a glass substrate covered by indium tin oxide (ITO) at 4000 r.p.m for 60s. During the spinning, chlorobenzene (anti-solvent) was dropped on the substrate to control the morphology of the film. The samples were then annealed at 70$_o$C for 20 min in a nitrogen-filled glovebox.

*Optical Measurements:* The time-integrated PL spectra were excited using second harmonic generation (3.1 eV) or the fundamental harmonic (1.55 eV) of a mode-locked Ti: sapphire laser (Mira 900, Coherent). The typical temporal pulse width was around 150 fs, with a repetition rate of 76 MHz. The laser power was adjusted using neutral density filters during the measurement. The excitation beam was spatially limited by an iris and focused with a 150-mm focal length lens. PL was collected by a spectrometer and recorded by an Imaging EM CCD camera from Hamamatsu (Hamamatsu, Japan). The time resolved PL spectra were dispersed by an imaging spectrometer and detected using a Hamamatsu streak camera. Depending on the time-window used, the time resolution varied. When the streak camera is working in Synchroscan mode, the time resolution is around 10 ps for a 2 ns time window. In single sweep mode, the time resolution is around 1% of the time window, and a pulse picker is used to reduce the repetition rate of the exciting pulses.

*Other Characterization:* Powder X-ray diffraction data were collected using a Bruker D8 Advance diffractometer in Bragg-Brentano geometry and operating with Cu Kα radiation. Single crystal X-ray diffraction was performed using a Bruker D8 Venture diffractometer operating with Mo Kα radiation and equipped with a Triumph monochromator and a Photon100 area detector. The sample was mounted in a nylon loop using cryo-oil and cooled using a nitrogen flow from an Oxford Cryosystems Cryostream Plus. The data were processed using the Bruker Apex II software. The structure was solved and refined using the SHELXTL software. The PL mapping images were captured by a fluorescence microscope under a defocused, spatially homogeneous 488 nm continuous wave laser beam excitation. Topography characterization of the sample was performed using AFM (VEECO) in tapping mode.

**Acknowledgements**

The authors would like to thank Arjen Kamp and Theodor Zaharia for their technical support. H.H. Fang and M.A. Loi acknowledge the financial support of the European Research Council (ERC Starting Grant "Hy-SPOD" No. 306983). S. Adjokatse and M.E. Kamminga acknowledge financial support from the NWO Graduate Programme 2013 (no. 022.005.006). The authors declare no competing financial interest.

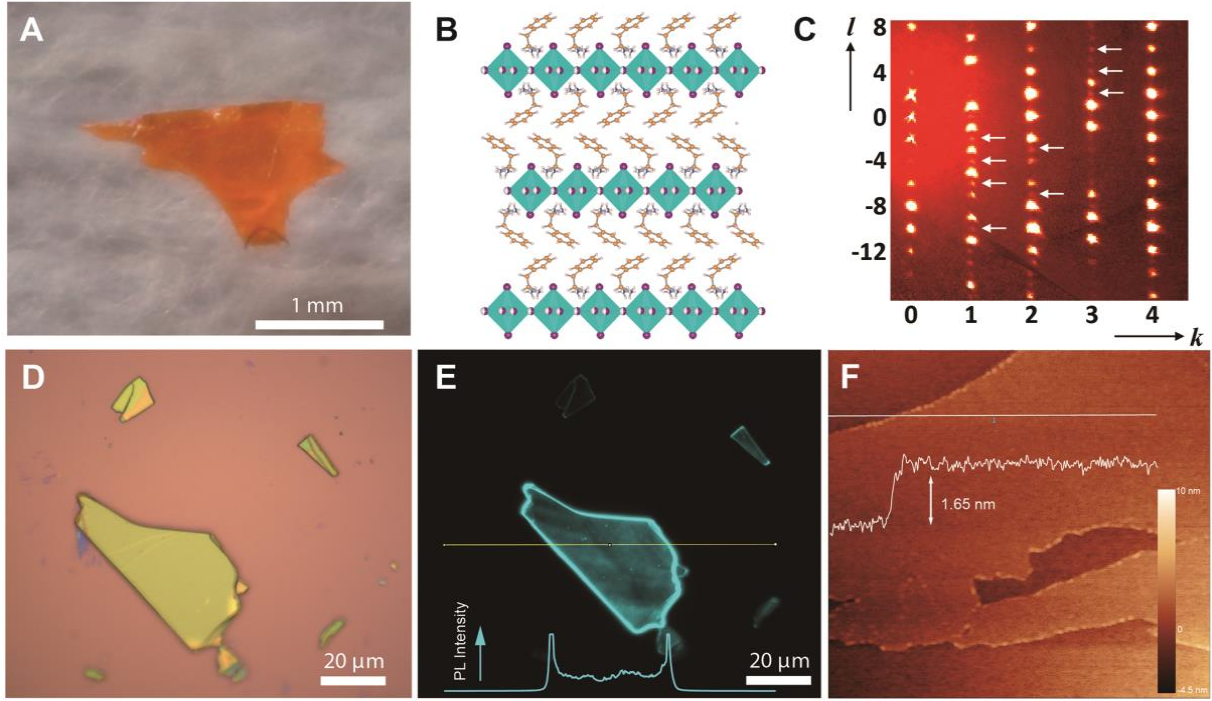

**Figure 1. Characterization of layered hybrid perovskite crystals.** (**A**) A photograph of a single crystal of (PEA)$_2$PbI$_4$. (**B**) Crystal structure of (PEA)$_2$PbI$_4$ at 100 K. (**C**) 1*kl* reciprocal lattice plane reconstructed from raw single crystal diffraction data of (PEA)$_2$PbI$_4$ at 100 K. White arrows indicate spots that violate the *hkl*, $k+l=2n$ condition for lattice centering. (**D**) Optical microscope image and (**E**) fluorescence microscope image of (PEA)$_2$PbI$_4$ crystal prepared by mechanical exfoliation. The inset of (E) shows the PL intensity profile. (**F**) AFM images of the surface of a (PEA)$_2$PbI$_4$ crystal, showing step-like morphology.

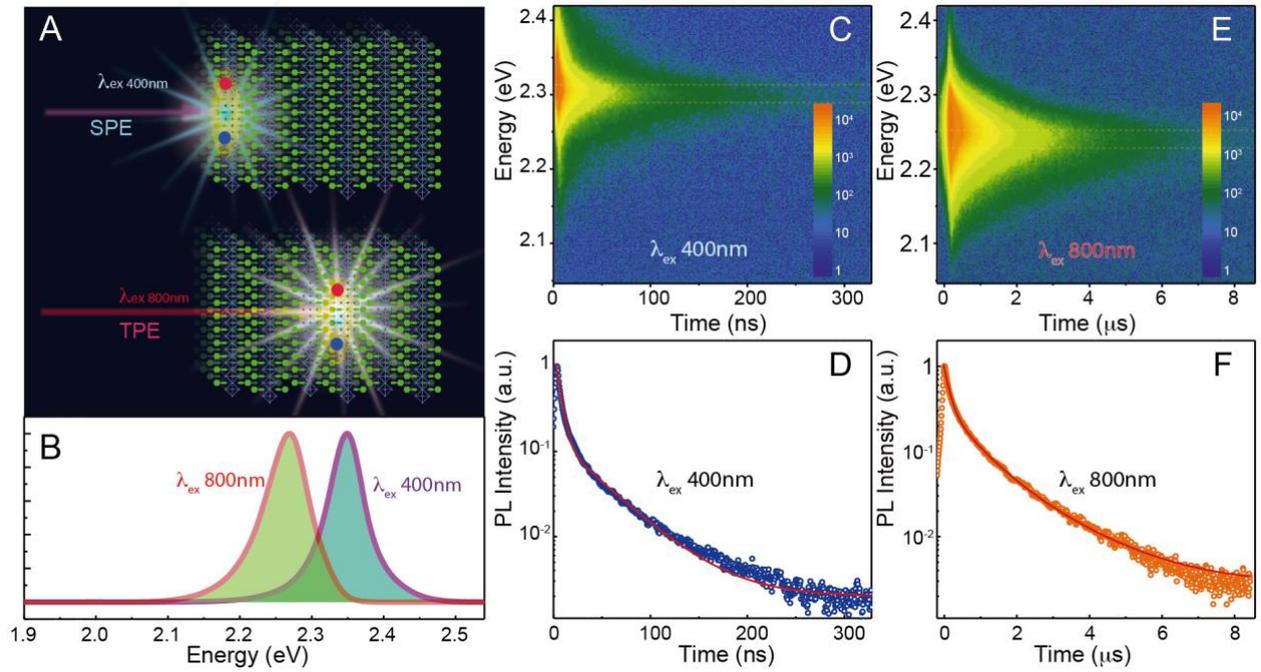

**Figure 2. One- and two-photon excited photoluminescence from layered perovskite single crystals.** (**A**) Schematic of the (PEA)₂PbI₄ crystal under single-photon (3.1 e) and two-photon (1.55 eV) excitation. (**B**) PL spectra of (PEA)₂PbI₄ crystal under single- and two-photon excitation. (**C**) Spectral- and time-resolved PL images of (PEA)₂PbI₄ crystal under excitation of 3.1 eV (400 nm), and (**D**) corresponding normalized PL decay of the emission peak. (**E**) Spectral- and time-resolved PL images of (PEA)₂PbI₄ single crystal under two-photon excitation, and (**F**) corresponding normalized PL decay of the emission peak.

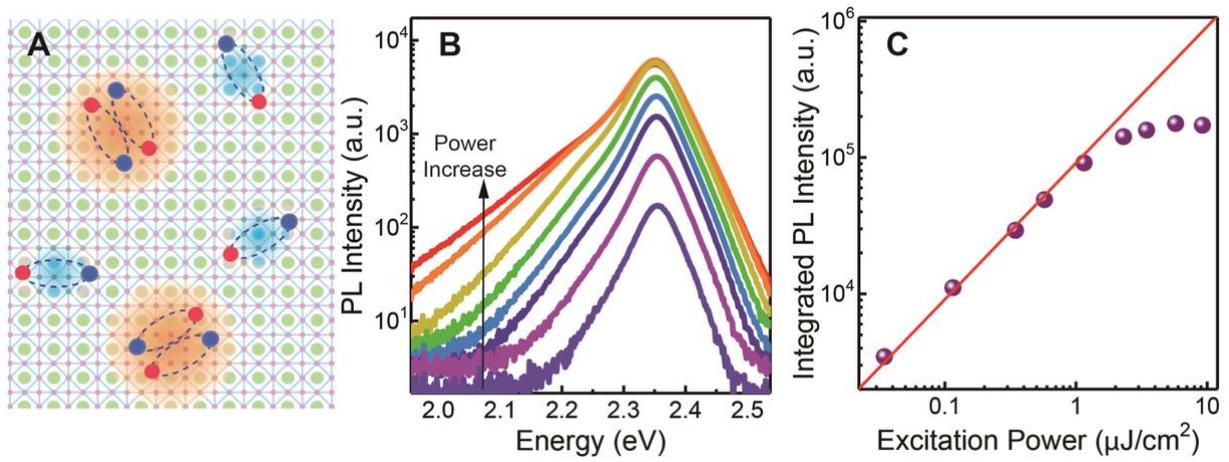

**Figure 3. Room temperature biexciton emission in layered perovskite crystal.** (**A**) Schematic of the formation of biexcitons in layered perovskite. (**B**) Power-dependent photoluminescence at room temperature showing the emergence of an additional PL shoulder at high-power excitation. (**C**) Integrated PL intensity as a function of excitation power. The slope is 1 at low excitation power.

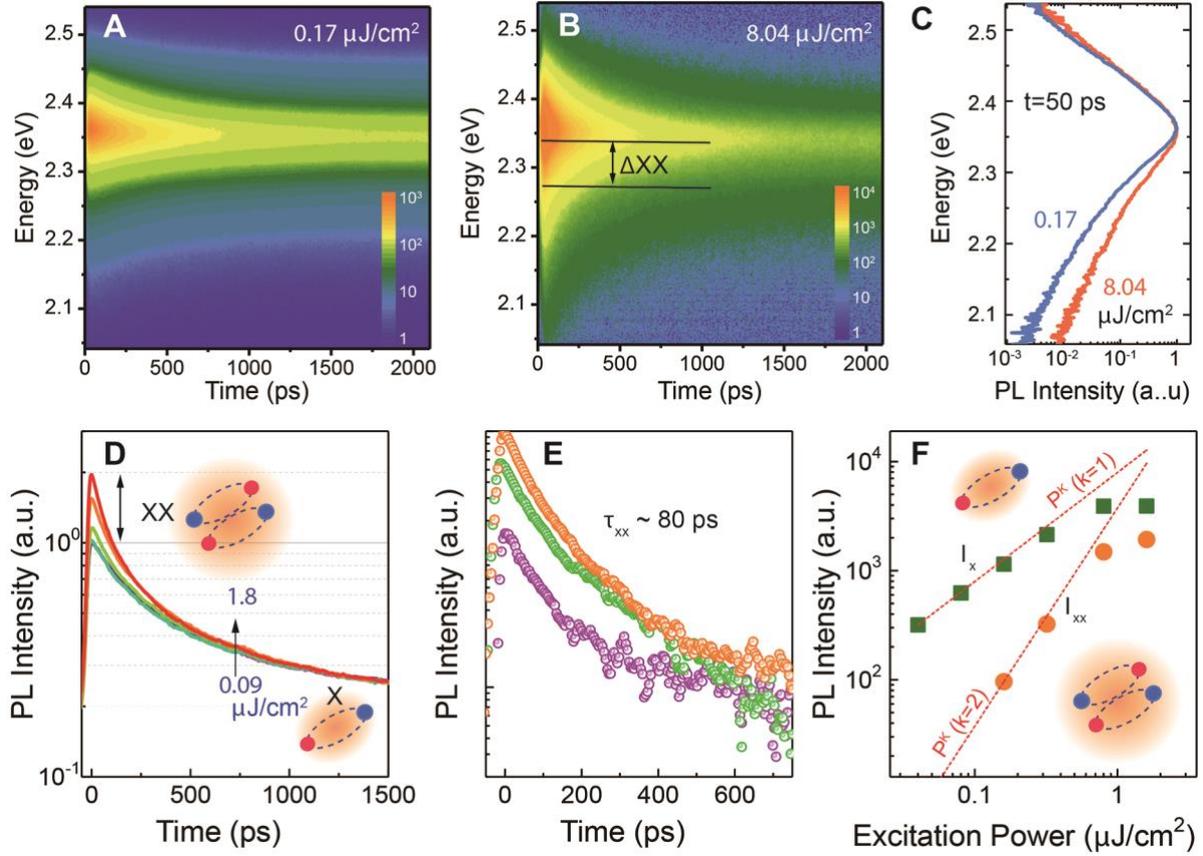

**Figure 4. Room temperature time-resolved photoluminescence from (PEA)$_2$PbI$_4$ crystal.** Spectral- and time-resolved PL images of (PEA)$_2$PbI$_4$ crystal, showing (**A**) low-fluence (0.17 μJcm$^2$) and (**B**) high-fluence (8.04 μJ/cm$^2$) emission after excitation at 400nm. (**C**) Comparison of PL spectra taken at t = 50 ps for high and low excitation density, showing additional emission at a low energy shoulder. (**D**) Excitation power-dependent PL dynamics of (PEA)$_2$PbI$_4$ crystal. The PL decays are normalized to match the late-time tails. The early time short-lived PL components at high excitation power show biexciton recombination. (**E**) Isolated biexciton recombination under three different excitation fluences (from 0.36 to 1.8 μJ/cm$^2$). The biexcitonic components are calculated from the single-exciton decay by subtracting the dynamics measured at low pump fluence (0.09 μJ/cm$^2$). (**F**) Excitation power-dependence of the PL amplitude at t = 0, and the amplitude of the biexcitonic component extracted from the fast relaxation contribution, confirming the room temperature biexciton emission.

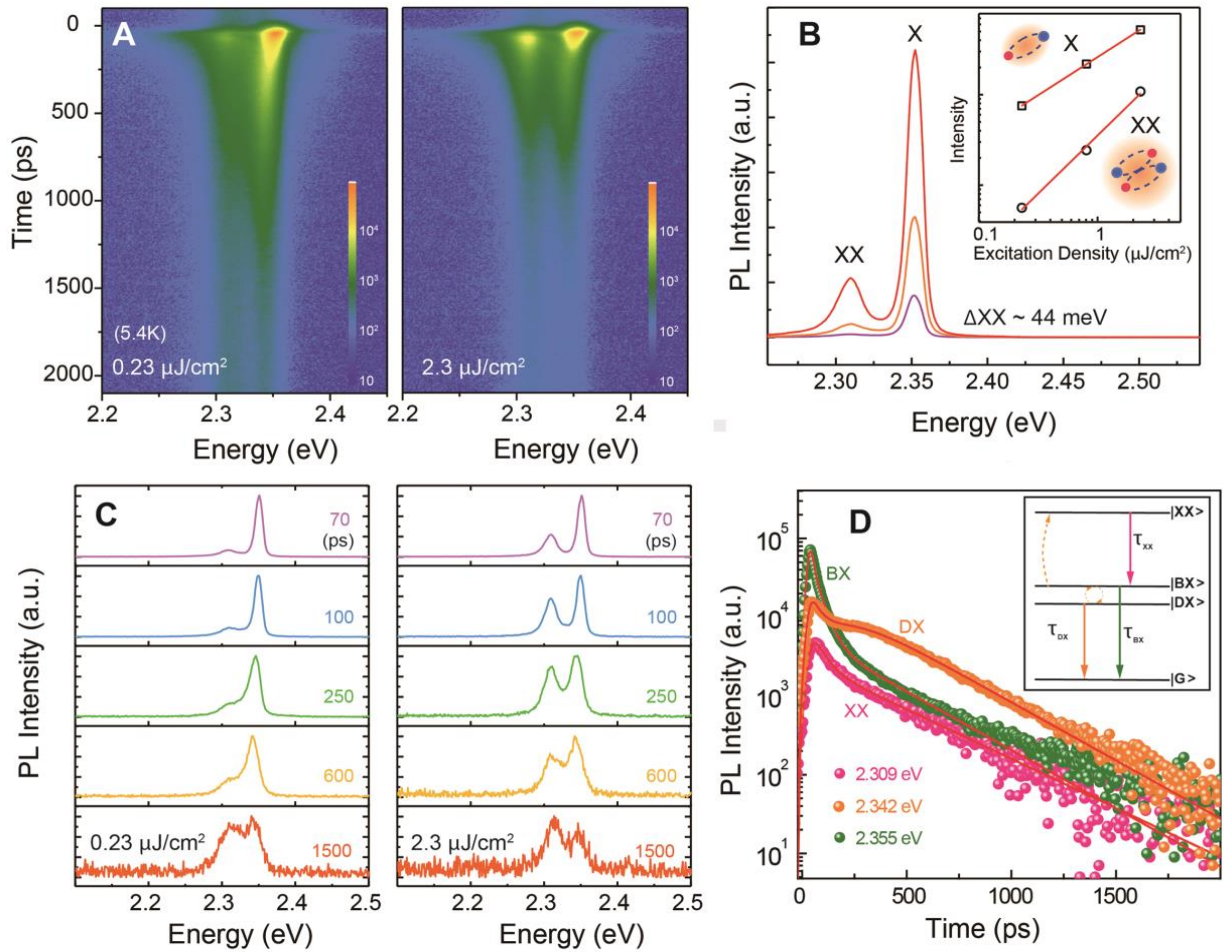

**Figure 5. Fine excitonic emission at 5.4 K in (PEA)$_2$PbI$_4$ single crystals.** (**A**) Spectral- and time-resolved PL images of (PEA)$_2$PbI$_4$ crystal measured at 5.4 K for two excitation densities at 400nm. (**B**) Power-dependent PL spectra of (PEA)$_2$PbI$_4$ crystal at t = 70 ps. The inset shows the intensity of the single exciton (X) and biexciton (XX) emission lines as a function of excitation density. (**C**) PL spectra, corresponding to different times after the excitation pulse (labelled in ps) showing the time-evolution of single exciton (X) and biexciton (XX) emission. The spectra are normalized to the maximum value. (**D**) Decay of the PL at 2.309 eV, 2.342 eV and 2.355 eV on a semilogarithmic scale. Inset: Proposed energy level diagram for (PEA)$_2$PbI$_4$ involving dark (DX) and bright (BX) exciton levels, as well as a biexciton level (XX).

# Supporting Information

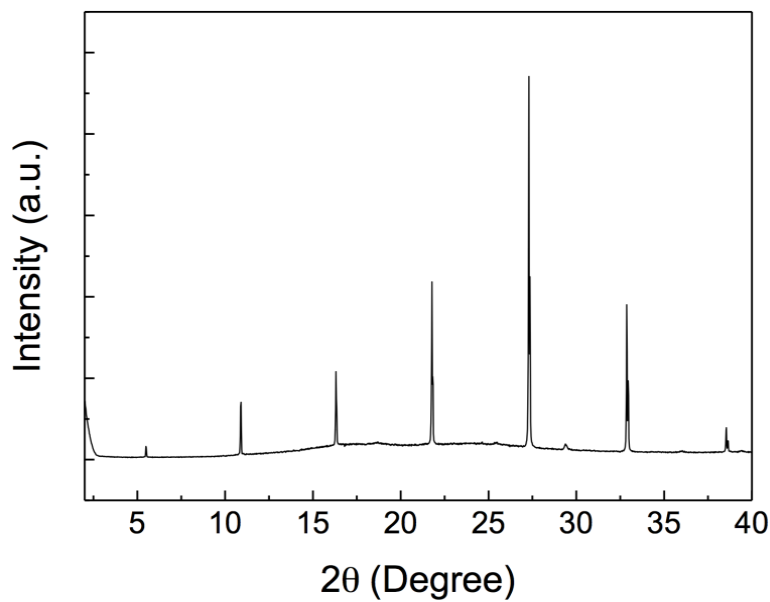

**Fig. S1**. X-ray diffraction data from a flake obtained from a mechanically exfoliated crystal.

**Fig. S2.** Example of a reciprocal lattice plane reconstructed from raw XRD data at 100 K, showing streaked intensity along the reciprocal [001]* direction at half-integer $h$ and $k$ indices, attributed to local ordering of octahedral rotations.

Inspection of reciprocal space plots reconstructed from the raw XRD data (an example of which is shown in Fig. S2) revealed streaks of intensity in the reciprocal [001]* direction at half-integer values of $h$ and $k$. These streaks indicate that the unit cell is locally enlarged in the $ab$-plane by $\sqrt{2}a$ x $\sqrt{2}b$ to form a pseudo C-centered lattice with double the unit cell volume, but with a very short coherence length. It is likely that the octahedra are rotated in ordered fashion in the ab-plane, but that the correlation length of this order is short. Therefore, the structure was solved using disordered iodine positions in the *ab*-plane, which correspond to two possible rotation angles of the PbI$_6$ octahedra (see *ab*-plane view below). The NH$_3$ group of the PEA molecule is also disordered over two positions; this disorder is most likely induced by hydrogen bonding with the disordered octahedra.

Structures were also determined from XRD data collected at 200 K and 295 K. No major changes were observed except for larger ADPs (atomic displacement parameters) on the PEA molecule, indicating an increasing degree of disorder with temperature. We conclude that there is no phase transition in the temperature range 100 – 295 K.

**Table S1.** The unit cell parameters of (PEA)$_2$PbI$_4$ at 100, 200 and 295 K.

| Temperature | *a* (Å)    | *b* (Å)    | *c* (Å)     | β (degrees) |
|-------------|------------|------------|-------------|-------------|
| 100 K       | 6.1594(4)  | 6.0991(4)  | 32.261(2)   | 94.299(2)   |
| 200 K       | 6.1845(9)  | 6.1305(8)  | 32.501(4)   | 94.064(6)   |
| 295 K       | 6.262(4)   | 6.196(4)   | 33.334(17)  | 92.50(2)    |

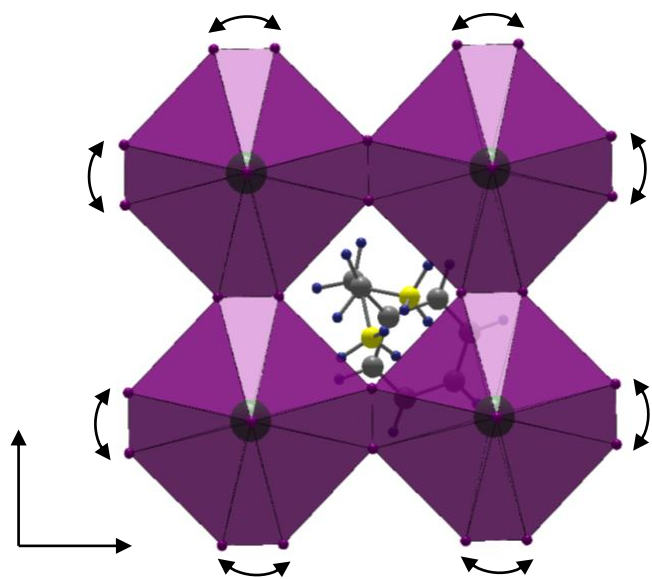

**Fig. S3.** *ab*-plane representation of (PEA)$_2$PbI$_4$ structure. The arrows indicate two possible rotation angles of the octahedra; these rotations are ordered over very short length-scales. The two disordered positions of the NH$_3$ group of the PEA molecule are also shown.

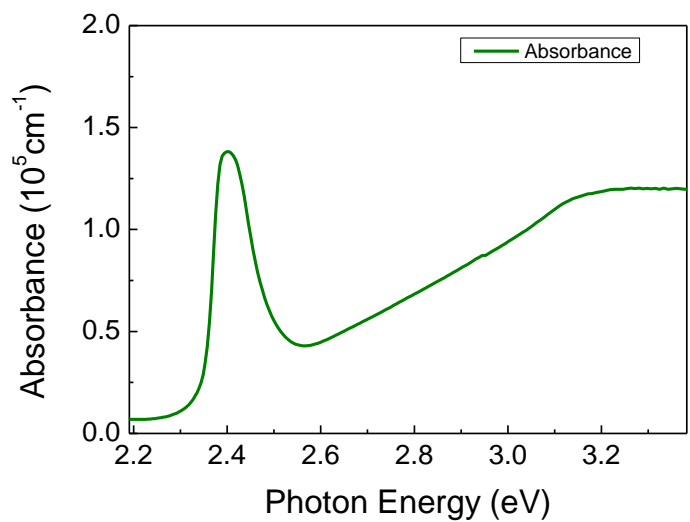

**Fig. S4**. UV-vis. spectrum of (PEA)$_2$PbI$_4$ thin film at room temperature.

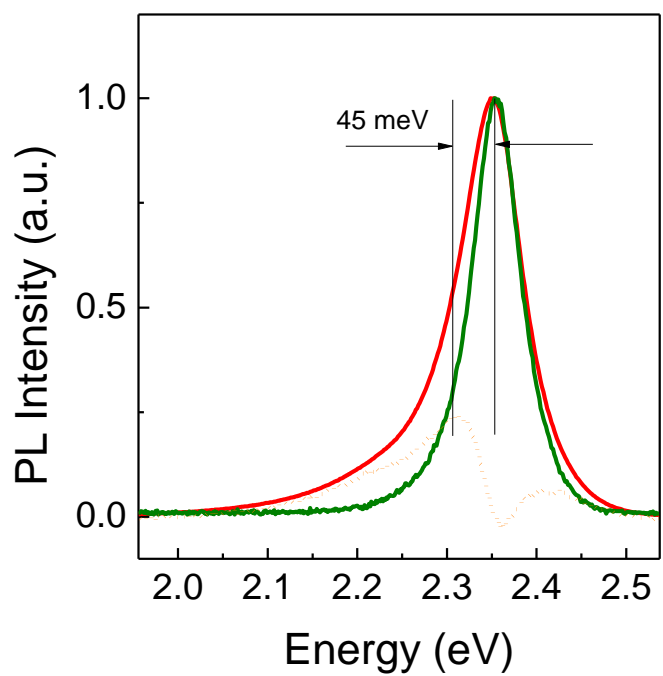

**Fig. S5**. Room temperature PL Spectra under excitation of 400 nm laser at low power density (olive) and high excitation (red). The dashed orange line shows the difference spectrum between low and high excitation.

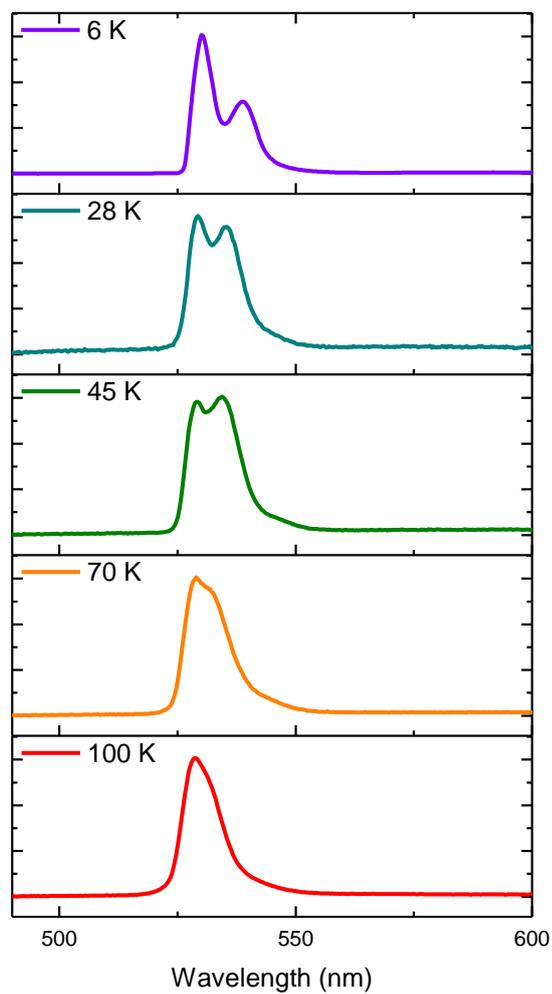

**Fig. S6.** PL spectra of (PEA)$_2$PbI$_4$ at different temperatures.

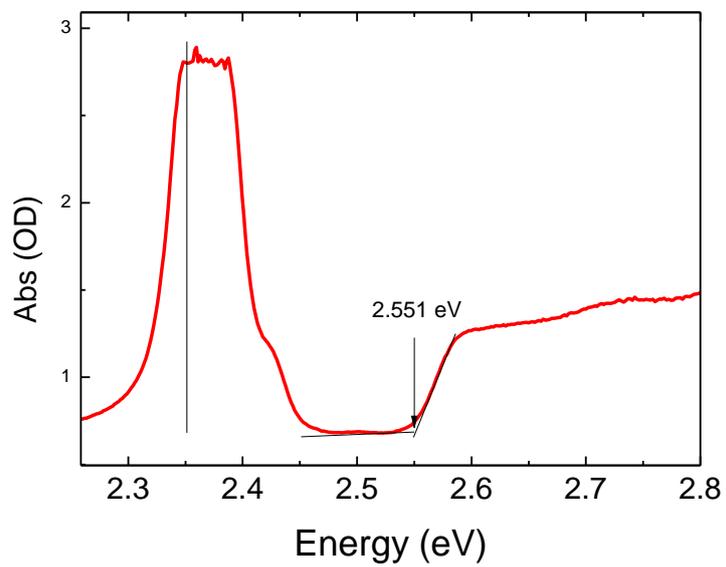

**Fig. S7**. UV-vis. spectrum of (PEA)$_2$PbI$_4$ thin film at 5.4 K.

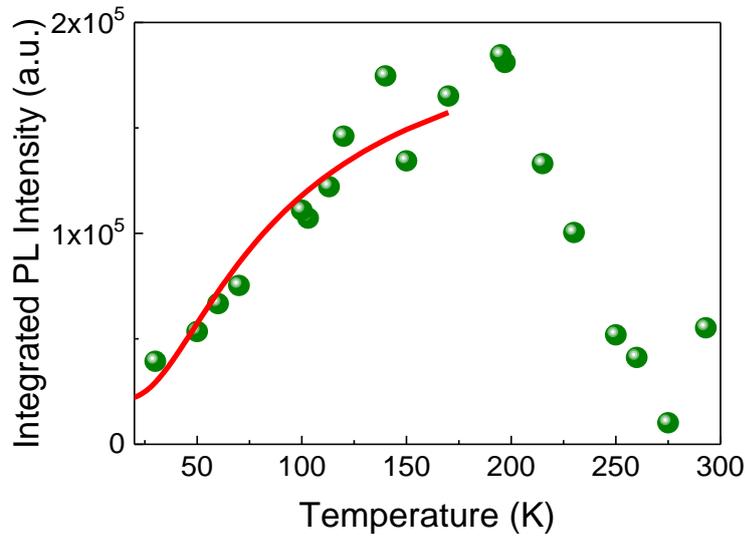

**Fig. S8.** Integrated intensity of luminescence band in (PEA)$_2$PbI$_4$ as a function of temperature. At low temperatures (30- 150 K) the intensity decreases when temperature decreases, while at temperatures higher than 200 K, the luminescence intensity increases when cooling down. The temperature dependence of the PL intensity at low temperature can be fitted with a two-level model, including a high energy exciton state which is bright and a lower energy state, with an energy separation $\Delta E$. Assuming a thermal equilibrium between them, the PL intensity is simply proportional to: $\exp(-\Delta E/k_B T)/[\exp(-\Delta E/k_B T)+1]$, where $\Delta E$ is the bright-dark exciton energy difference.[1] The red curve is obtained from this model with the parameter $\Delta E = 10$ meV.

[1] X.-X. Zhang, Y. You, S. Y. F. Zhao, T. F. Heinz, *Phys. Rev. Lett.* **2015**, *115*, 257403.